\documentclass[aps,twocolumn,pra,showpacs,floatfix]{revtex4}

\usepackage{dcolumn}

\begin{document}

\title{\bf Electric dipole moments of
actinide atoms and RaO molecule}
\author{V. V. Flambaum}
\affiliation{School of Physics, University of New South Wales,
Sydney 2052, Australia and Institute for Advanced Study,
Massey University (Albany Campus), Private Bag 102904, North Shore MSC
 Auckland, New Zealand}

\date{\today}

\begin{abstract}

We have calculated the atomic electric dipole moments (EDMs)
induced in $^{229}$Pa and $^{225}$Ac by their respective nuclear 
Schiff moments $S$. The results are
$d(^{229}{\rm Pa})= - 9.5 \cdot 10^{-17} [S/(e \cdot fm)] e \cdot cm=
 - 1.1 \cdot 10^{-20} \eta\,\, e \, cm\,$;
$d(^{225}{\rm Ac})=- 8.6 \cdot 10^{-17} [S/(e \, fm)] e \, cm=
 - 0.8 \cdot 10^{-21} \eta\,\, e \, cm\,$. EDM of $^{229}{\rm Pa}$
is $3 \cdot 10^4$ times larger than $^{199}$Hg  EDM and 40 times
 larger than $^{225}$Ra EDM. Possible use of actinides in solid
state experiments is also discussed.
 The T,P-odd spin-axis interaction in RaO molecule
is 500 times larger than in TlF. 
\end{abstract}


\maketitle

\section{Introduction}

Measurements of atomic EDM allows one to test CP violation theories 
beyond the Standard Model (see, e.g.~\cite{Dmitriev,Ginges}).
The best limits on atomic EDM have been obtained for diamagnetic
atoms  Hg~\cite{Romalis} and
Xe~\cite{Rosenberry}; there is also limit on T,P-odd spin-axis interaction in
 TlF molecule\cite{tlfedm}. EDM of diamagnetic atoms
 and molecules is induced by the
interaction of atomic electrons with the nuclear
Schiff moment. Schiff moments produced by the  nuclear T,P-odd interactions
have been calculated in Refs. \cite{sushkov84,sushkov86,flambaum02,DS03,DS04,
DF05,dejesus05}.
2-3 orders of magnitude enhancement
can exist in  nuclei with octupole deformation
\cite{Spevak} or soft octupole vibrations \cite{engel00}.
This motivated new generation of atomic experiments with $^{225}$Ra
(see e.g. \cite{Guest}) and $^{223}$Rn. Current status of EDM experiments
 and theory can be found on the website of the INT workshop \cite{INT}. 
Most accurate calculations of  atomic EDM produced by the nuclear Schiff
 moments have been performed for  Hg, Xe, Rn, Ra,Pu \cite{Kozlov},
 Yb and He \cite{Dzuba07} atoms and TlF molecule \cite{Kozlov,petrov02}.
In this work  we would like to note that several orders of magnitude
larger effects appear in different systems.
\section{${\rm RaO}$ molecule}
% Let us start from RaO molecule.
 Due to the octupole enhancement \cite{Spevak} the Schiff
moments of $^{225,223}$Ra exceeds that of $^{205,203}$Tl
 \cite{sushkov84,sushkov86}
 $\sim 200$ times. There is an additional
 enhancement due to the higher nuclear charge of Ra since the effect produced
by the Schiff moment increases faster than $Z^2$. 
 Therefore , one may
hope that experiments with RaO may be up to 3 orders of magnitude
more sensitive to the nuclear T,P-violating interactions  than the experiments
with TlF. The results of molecular calculations are usually
 expressed in terms of the following matrix element:
\begin{eqnarray}
\label{tlf1}
  X &=& -\frac{2 \pi}{3}
  \langle \Psi_0|[\mbox{\boldmath$\nabla$} \cdot {\bf n},\delta({\bf R})]
  |\Psi_0 \rangle,
\end{eqnarray}
where $\Psi_0$ is the ground state wave function and $\bf n$ is the unit
vector along the molecular axis. The T,P-odd spin-axis interaction
constant can be  expressed in terms of the Schiff moment $S$
 (see e.g. \cite{Kozlov}):
\begin{eqnarray}
\label{tlf4}
  \langle \Psi_0| H_W |\Psi_0 \rangle &=& 6\, X
  {\bf S}\cdot{\bf n}.
\end{eqnarray}
Here  ${\bf S}$ is the Schiff moment vector.
For TlF molecule $X=7475$ atomic units (a.u.) \cite{Kozlov,petrov02}.
We use this result to estimate $X$ for RaO. One can view TlF
molecule as an ion compound Tl$^+$F$^-$. The electronic configuration
of   Tl$^+$ is $...6s^2$.
 The electric field
of F$^-$ produces s-p hybridization of Tl$^+$ orbitals
and non-zero matrix element $X$ in Eq.(\ref{tlf1}).
The electronic configuration of Ra,  $...7s^2$, is similar to 
Tl$^+$, and the T,P-odd spin-axis
 interaction in RaO is also due to the s-p hybridization.
A simplest way to estimate $X$ for RaO is to use known
$Z$ dependence of the Schiff moment effect: $Z^2 R(Z \alpha)$ where
  $ R(Z \alpha)$ is the relativistic factor \cite{sushkov84}.
This gives $X(RaO)/X(TlF) \approx 1.6 S(Ra)/S(Tl)$.
 A slightly more accurate result may be obtained using existing
atomic EDM calculations (atomic EDM and molecular T,P-odd spin-axis
 interaction depend on the same
matrix elements of the Schiff moment field).
 Hg atom has the same electronic configuration as Tl$^+$. The ratio of
Ra and Hg EDM was calculated in \cite{Kozlov}: 
$d(Ra)/d(Hg)=3.04 S(Ra)/S(Hg)$. The larger value for Ra is
due to higher nuclear charge: $Z$=88 for Ra, $Z$=80 for Hg and  $Z$=81 for Tl. 
Using $d(Ra)/d(Hg)=3.04 S(Ra)/S(Hg)$ we obtain  an estimate
$X(RaO)/X(TlF) \approx  2.8 S(Ra)/S(Tl)$. As a final value we will use
an intermediate result $X(RaO)/X(TlF) \approx  2.2 S(Ra)/S(Tl)$
which is between the EDM estimate and the relativistic factor estimate.
 This gives the T,P-odd spin-axis
 interaction in RaO 
\begin{eqnarray}
\label{tlf7}
  \langle \Psi_0| H_W |\Psi_0 \rangle &=& 1. \times 10^5
  ({\bf S}\cdot {\bf n})\,\, \mbox{a.u.}\ .
\end{eqnarray} 
%The accuracy of this estimate is better than factor of 2. 
The Ra Schiff moment $S$ is 200 times larger than the Tl Schiff moment,
altogether we obtain 500 times enhancement in RaO in comparison with TlF.
Note that the error of this number is probably dominated by the nuclear
 calculations of the Schiff moments.

\section{EDM of actinide atoms}
  The largest Schiff moment was found for $^{229}$Pa \cite{Spevak}
where the atomic calculation of EDM is absent. Below we obtain
the result for this EDM.
 For the first time the enhancement of P,T-violation
in  $^{229}$Pa nucleus was found by Haxton and Henley 
 in Ref. \cite{Haxton}. This nucleus contains very close
 excited level (220 eV) which has the same spin as the ground
state level ($I$=5/2) and  opposite
parity. These ground and excited states, $5/2^+$ and $5/2^-$,  
can be mixed by the nucleon P,T-odd interaction. Haxton and Henley
performed calculations in the Nilsson model (using single-particle
orbitals for the quadrupole nuclear deformation)
and found that nuclear EDM and magnetic quadrupole moment 
 are significantly
enhanced.  Unfortunately, they did not calculate Schiff moment.
Calculation of the Schiff moment was performed in Ref. \cite{Spevak}
assuming different model (octupole nuclear deformation)
which gives an additional enhancement due to the collective
nature of  the Schiff moment in  nuclei with octupole deformation.
A similar mechanism produces enhancement of the T,P-odd electric octupole
moment \cite{Murray}. It is interesting that in  $^{229}$Pa
all four T,P-odd nuclear moments (Schiff, EDM, magnetic quadrupole
 and octupole)
 contribute to atomic EDM. Let us start from the Schiff moment which gives
a dominating contribution in  $^{229}$Pa.

 The electron configuration of Pa is $...7s^2 5f^2 6d$.
The Schiff moment field  is confined inside the nucleus.  
The high-wave $6d$ and $5f$ electrons practically do not penetrate
inside the nucleus and
 have very small matrix elements for Schiff moment field.
If we neglect these small matrix elements, the atomic EDM
comes from the  Ra-like core $...7s^2$. 
 In this approximation we may use the result for Ra,
$d= - 8.23 \cdot 10^{-17} [S/(e \, fm)] e \, cm$ from Ref. \cite{Kozlov},
to calculate Pa EDM. The coefficient actually should be slightly
larger since the Pa charge $Z=91$ is larger than the Ra charge $Z=88$.
Another reference point is Pu, Z=94,
where $d=- 10.9 \cdot 10^{-17} [S/(e \cdot fm)] e \cdot cm$ from
 Ref. \cite{Kozlov}. Pu has the electron configuration   $...7s^2 5f^6$
where the contribution of $5f$ electrons is  not very important
 (as explained above).
The Pa atom,  $Z=91$, is exactly in between   Ra, $Z=88$, and  Pu, Z=94.
Therefore,  we take the average value as  Pa EDM,
 $d= - 9.5 \cdot 10^{-17} [S/(e \cdot fm)] e \cdot cm$.
The accuracy of this result is about 20 \% (see  Ref. \cite{Kozlov}).

 Now discuss the contributions of other T,P-odd moments.
 Nuclear EDM contributes in combination with magnetic hyperfine
interaction between nucleus and atomic electrons \cite{Schiff}.
However, this contribution has relatively slow increase with
nuclear charge, $\sim Z$, and may be neglected. The  contributions
of magnetic quadrupole, electric octupole and Schiff moments
increase faster than $Z^2$.  Electric octupole and magnetic quadrupole
induce atomic EDM only if electron angular momentum $J$ is not zero 
(since the EDM vector $d_i$ can only be produce from nuclear
 magnetic quadrupole
tensor $M_{ik}$ as $d_i \sim M_{ik}J_k$ or octupole third rank tensor 
as $d_i \sim O_{ikj}J_k J_j$). The electron angular momentum $J=11/2$
 is actually carried out by $6d$ and $5f$ electrons. The matrix elements of
very singular magnetic quadrupole and electric octupole fields for these
orbitals are small since these matrix element comes
from small distances where the high-wave electrons do not penetrate.
If we neglect these small matrix elements, the atomic EDM
comes from the  Ra-like core $...7s^2$ which has zero electron angular
momentum and no contributions from the magnetic quadrupole and electric
 octupole. Atomic EDM in this approximation
 comes entirely from the Schiff moment field which mixes $s-p$ orbitals. 
There are additional arguments why we do not need to include
the electric octupole and magnetic quadrupole contributions into our
 approximate calculations.
Without any enhancement (e.g. for spherical nuclei), the electric
octupole contribution to atomic EDM is substantially smaller
than the magnetic quadrupole and Schiff contributions   
(see comparison of the corresponding matrix elements in  Ref. \cite{Murray}).
 The octupole deformation (or the soft octupole mode)
 gives the collective enhancement of the Schiff and octupole moments,
however, it does not enhance the magnetic quadrupole
(the small nuclear energy denominator is a common factor
for all three contributions, so it does not influence the ratio
of them). These arguments stress again importance of the Schiff moment
contribution.

The Schiff moment of $^{229}$Pa was calculated in Ref. \cite{Spevak}:
$S=1.2 \cdot 10^{-4}$ e\, fm$^3$ $\eta$ where $\eta$ is the dimensionless
strength of the nucleon $P,T-odd$ interaction in units of the Fermi constant. 
Substituting this value we obtain EDM of $^{229}$Pa atom:
\begin{equation}
\label{PaEDM}
d= - 1.1 \cdot 10^{-20} \eta\,\, e \, cm\,.
\end{equation}
This value is $3 \cdot 10^4$ times larger than $^{199}$Hg atomic EDM,
$4 \cdot 10^{-25}\eta\,\, e \, cm\,$, and 40 times larger than
  $^{225}$Ra EDM, $2.6  \cdot 10^{-22}\eta\,\, e \, cm\,$
  (this comparison is  based on the
  Schiff moments from Refs. \cite{Spevak,sushkov86} and the atomic calculations
for Hg and Ra from Ref.  \cite{Kozlov} ).

A similar calculation for  $^{225}$Ac ($Z=89$, atomic configuration
$...7s^2 6d$, $J=3/2$) gives 
\begin{equation}
\label{AcEDM}
d= - 8.6 \cdot 10^{-17} [S/(e \, fm)] e \, cm=
 - 0.8 \cdot 10^{-21} \eta\,\, e \, cm\,.
\end{equation}
 
 We would like to suggest another possible application of actinides.
  Recently, the measurements of electron EDM and nuclear Schiff moments in 
the  solid compounds contaning  rare-earth atoms (e.g. gadolinium)
  have been proposed
 \cite{Lam,Hunter,Sushkov}
(corresponding calculations have been performed in \cite{Sushkov1,Sushkov}).
The actinides are electronic analogues of rare-earth atoms.
 Because of  rapid increase of atomic EDM with nuclear
charge it may be worth considering similar compounds with  actinides.
For example, uranium and thorium have isotopes which are practically
stable. Atomic EDM induced by the electron EDM increases with $Z$ as
\cite{Flambaum}
\begin{equation}
d \sim \frac{Z^3}{(j+1)(4 \gamma^2 -1)\gamma} 
\end{equation} 
where $\gamma=((j+1/2)^2-Z\alpha^2)^{1/2}$ and $j$ is the electron angular
momentum (maximal contribution comes from $j=1/2$).
 Comparing uranium ($Z=92$) with gadolinium  ($Z=64$) we see that
in uranium compounds the effect is  $\sim$5 times larger. A similar enhancement
happens for the effects induced by the nuclear T,P-odd moments.

%\section*{Acknowledgments}

This work was supported in part by the Australian Research Council.

\end{document}